\documentclass[aps,pra,showpacs,twocolumn,superscriptaddress]{revtex4}
\usepackage{graphicx}
\usepackage{dcolumn}
\usepackage{bm}
\newcommand{\ket}[1]{|#1\rangle}

\begin{document}

\title{Direct measurement of the quantum state of the electromagnetic field in a superconducting transmission line}

\author{F. de Melo}
\affiliation{Instituto de F\'{\i}sica, Universidade Federal do Rio
de Janeiro, Cx. P. 68528,\\
21941-972 Rio de Janeiro, RJ, Brazil}

\author{L. Aolita}
\affiliation{Instituto de F\'{\i}sica, Universidade Federal do Rio
de Janeiro, Cx. P. 68528,\\
21941-972 Rio de Janeiro, RJ, Brazil}

\author{F. Toscano}
\affiliation{Instituto de F\'{\i}sica, Universidade Federal do Rio
de Janeiro, Cx. P. 68528,\\
21941-972 Rio de Janeiro, RJ, Brazil}
\affiliation{Funda\c c\~ao
Centro de Ci\^encias e Educa\c c\~ao Superior \`a Dist\^ancia do
Estado do Rio de Janeiro, 20943-001 Rio de Janeiro, RJ, Brazil}

\author{L. Davidovich}
\affiliation{Instituto de F\'{\i}sica, Universidade Federal do Rio
de Janeiro, Cx. P. 68528,\\
21941-972 Rio de Janeiro, RJ, Brazil}

\date{\today}
\begin{abstract}

We propose an experimental procedure to directly measure the state
of an electromagnetic field inside a resonator, corresponding to a
superconducting transmission line, coupled to a Cooper-pair box
(CPB). The measurement protocol is based on the use of a dispersive
interaction between the field and the CPB, and the coupling to an
external classical field that is tuned to resonance with either the
field or the CPB. We present a numerical simulation that
demonstrates the feasibility of this protocol, which is within reach
of present technology
\end{abstract}
 \maketitle

Superconducting electrical devices have been experimentally proven
to be  serious candidates for the realization of quantum information
processing tasks~\cite{makhlin:357}. Coherent control and near
unit-visibility
Rabi-oscillations~\cite{wallraff:162,nakamura:246601}, coupling of
two CPB-qubits~\cite{pashkin:823} and the implementation of
conditional gates~\cite{yamamoto:941} are striking experiments that
demonstrate the high level of control achieved on these systems.
Furthermore, a scalable architecture for quantum computation has
already been envisioned~\cite{you:197902}.

On the other hand, recent demonstrations  of Jaynes-Cummings-like
dynamics between a CPB-qubit and the quantized mode of a
superconducting transmission line resonator (which acts as a
quasi-1D cavity)~\cite{wallraff:162,blais:062320} have shown that
many of the tools originally developed within the context of quantum
optics can now be extended to solid state physics. Once coherent
control and complete characterization of quantum states have been
achieved at the qubit level, it is natural to attempt such levels of
control for the electromagnetic field generated by the transmission
line. For its characterization one could, in principle, make use of
the well-known homodyne and heterodyne detection techniques. But,
since the field we would like to characterize is inside a resonator
and consists of a few photons, implementation of those techniques
turns out to be a non-trivial task. Homodyne detection has been
proposed for characterizing the state of the field leaking out from
a tridimensional cavity in \cite{santos:033813}, and for a
one-photon field leaking out of a 1-D cavity in \cite{solano}.
Nevertheless, it is very difficult to apply this procedure to
high-finesse cavities containing weak fields, since one would have
to distinguish a still weaker leaking field from the noise in the
detector. Furthermore, unavoidable absorption losses may lead to
poor reconstruction of the state of the intracavity field, as
pointed out in~\cite{khanbekyan:043807}.

To overcome these issues we propose here an experiment to directly
measure the Wigner function~\cite{wigner} of the electromagnetic
field inside a superconducting transmission line resonator coupled
to a CPB-qubit, via the measurement of the latter's populations. The
Wigner function contains all the information about the state of the
field, and is a useful tool for studying the decoherence-induced
quantum-to-classical transition, as it provides us with a
phase-space representation that can be compared to classical
probability distributions~\cite{toscano}. For a single mode of the
electromagnetic field, it is defined in terms of the respective
density operator $\hat{\rho}$ as~\cite{cahill}:
\begin{equation}
W(\alpha)=(1/\pi){\rm Tr}[\hat\rho\hat D(\alpha)\hat P \hat
D^{-1}(\alpha)]. \label{wigner}
\end{equation}
Here, $\hat D(\alpha)=e^{\alpha \hat a^\dagger -\alpha^* \hat a}$ is
the field displacement operator, which takes any coherent state
$\ket{\beta}$ to  $\ket{\beta+\alpha}$, up to a phase factor, and
$\hat P=e^{-i \pi \hat a^\dagger \hat a}$ is the parity operator,
which multiplies a Fock state $\ket{n}$ by a factor $(-1)^{n}$;
$\hat a$ and $\hat a^\dagger$ are respectively the photon
annihilation and creation operators of the mode. The displacement
operator can be operationally implemented, in a cavity QED (cQED)
setup \cite{raimond}, by injecting a coherent field with complex
amplitude $\alpha$ into the cavity. A protocol for the direct
measurement of the Wigner function was first proposed in
\cite{lutterbach:2547} and later experimentally carried out in
\cite{nogues:054101} for the microwave field inside a 3-D
high-quality-factor (Q) cavity. It involves injecting a microwave
field (complex amplitude $\alpha$) into the cavity, so as to
displace the field to be measured, and then sending an atom with two
of its levels, $|e\rangle$ and $|g\rangle$, interacting dispersively
with the displaced field. The atom is prepared in the state
$(|e\rangle+|g\rangle)/\sqrt{2}$, and, after leaving the cavity, is
submitted to a classical field, so that its state undergoes a
$\pi/2$ rotation [$|e\rangle\rightarrow (|e\rangle+|g\rangle)/2$,
$|g\rangle\rightarrow(-|e\rangle+|g\rangle)/2$]. Then the atomic
population is measured. The difference between the probabilities of
finding the atom in states $|e\rangle$ and $|g\rangle$ is
proportional to the value of the Wigner function of the cavity field
at the point $-\alpha$ in phase space.

It is not possible however to apply this protocol to the system here
considered, since in this case the atom (CPB-qubit) is always inside
the cavity and its interaction with the field cannot be turned off.
Nevertheless, we show here that it is still possible to directly
measure the Wigner function of the electromagnetic field in a
superconducting transmission line, via the Copper-pair box qubit.
Our method could also be applied to other systems involving the
interaction of a qubit with a resonator \cite{geller}.

\begin{figure}[t]
\centering
\begin{tabular}{cc}
\resizebox{!}{4.0cm}{\includegraphics{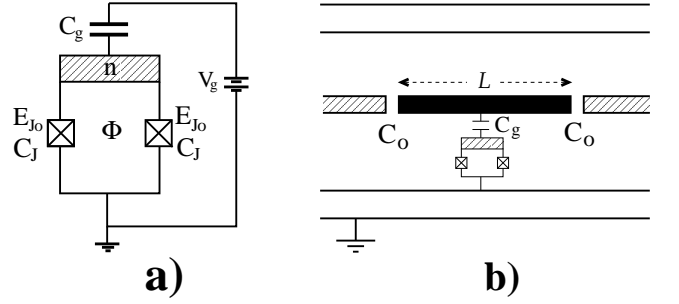}}
\end{tabular}
\caption{\footnotesize {\bf a)} \emph{The Cooper-pair box}: a
superconducting island with $n$ Cooper pairs in excess (dashed area)
with two Josephson Junctions (crossed boxes). {\bf b)}
\emph{Resonator-CPB-qubit composite system}: the central
superconducting waveguide (black) is coupled capacitively ($C_0$) to
other two superconducting lines (dashed), through which the
classical driving fields are pumped.} \label{sysfig}
\end{figure}

The system under consideration consists of a mesoscopic
superconducting island (see Fig.~\ref{sysfig}a) capacitively coupled
to the quantized field mode of a transmission line of length $L$
(see Fig.~\ref{sysfig}b). Details of this system can be found in
\cite{blais:062320}. The CPB Hamiltonian is given by
\cite{makhlin:357}:
\begin{equation}
\label{CPB-Hamiltonian} \hat{H}_{CPB}= 4 E_C(\hat n-n_g)^2- E_J \cos
\hat\Theta\,,
\end{equation}
where $\hat n$ is the number operator corresponding to the
Cooper-pair charges in excess on the island, and $\hat\Theta$ is the
average phase drop along the junctions ($\hat
n=-i\partial/\partial(\hat\Theta)$). Each junction is characterized
by a capacitance $C_J$ and a Josephson energy $E_{Jo}$. The
effective Josephson coupling $E_{J}=2E_{Jo}\cos(\pi\Phi/\Phi_o)$ can
be changed through an applied static magnetic flux $\Phi$
($\Phi_o=h/2e$ is the flux quantum and $e$ the electron charge). The
charging energy is $E_C = e^2/2C_{\Sigma}$ ($C_{\Sigma}=C_g+2C_J$)
and the gate charge is $n_g\equiv C_gV_g/2e$, which can be tuned by
the dc part of the potential gate $V_g$. The coupling to the
quantized field mode of the transmission line, of frequency
$\omega$, is taken into account through the quantum part of the gate
voltage, {\it i.e.} $V_g\equiv V_g^{\rm dc}+v$, where
$v=\sqrt{\hbar\omega/Lc}\,(\hat{a}^{\dagger}+\hat{a})$ ($c$ is the
transmission line capacitance per unit length), with $\hat a$ being
the annihilation operator for the transmission-line mode. Finally,
the Hamiltonian of the CPB-resonator system is obtained by adding,
to Eq.~(\ref{CPB-Hamiltonian}), the Hamiltonian of the oscillator
mode, {\it i.e.} $\hat{H}_{\rm
os}=\hbar\omega\hat{a}^{\dagger}\hat{a}$. In the charge regime, {\it
i.e} $\Delta_{\rm GAP} \gg E_C \gg 4\,E_J$ ($\Delta_{\rm GAP}$ is
the superconductor gap) the CPB can be treated as an effective
two-level system \cite{makhlin:357} of transition frequency
$\omega_0=\sqrt{E_J^2+[4E_C(1-2\,n^{\rm dc}_g)]^2}/\hbar$, with
$n^{\rm dc}_g\equiv C_gV^{\rm dc}_g/2e$. Within this regime, when
$n^{\rm dc}_g$ is around $1/2$ and in the rotating-wave
approximation, the Hamiltonian of the composite system reduces to
that of the Jaynes-Cummings (JC) model with coupling Rabi frequency
$g=(eC_g/C_{\Sigma})\sqrt{\hbar\omega/Lc}$. In order to drive the
composite system the transmission line is coupled capacitively
($C_o$) to an external classical microwave field, of frequency
$\omega_d$ and slowly varying complex amplitude
$\epsilon(t)=\epsilon_R(t)+i\epsilon_I(t)$, whose effect can be
modeled through the driving Hamiltonian
$\hat{H}_d=\hbar[\epsilon(t)\hat{a}^{\dagger}e^{-i\omega_d t}+
H.c.]$. A second-order perturbative calculation, in the dispersive
regime $|g\sqrt{\overline{n}+1}/\Delta|\ll 1$ ($\overline{n}$ is the
mean photon number and $\Delta\equiv\omega_0-\omega$ is the detuning
between the cavity and the CPB-qubit) yields, for the total system
dynamics (including the driving), in a reference frame rotating with
the driving field frequency, the effective Hamiltonian
\begin{equation}
\hat{H}^{int}_{eff}=\hbar(\omega-\omega_d)\hat{a}^{\dagger}\hat{a}
+\hbar\left[\epsilon(t)\hat{a}^{\dagger}+H.c.\right]
+(\hbar/2)\vec\Omega\cdot\vec{\hat\sigma} \,, \label{EffHamiltonian}
\end{equation}
where
\begin{eqnarray}
\vec\Omega&=&\Big[(2g/\Delta)\epsilon_R(t),
-(2g/\Delta)\epsilon_I(t),
\omega_0-\omega_d\nonumber\\
&+&(g^2/\Delta)(2\hat a^\dagger\hat a+1)\Big]\,,\label{n}
\end{eqnarray}
and $\vec{\hat\sigma}=(\hat\sigma_x, \hat\sigma_y, \hat\sigma_z)$
are the Pauli matrices.

Hamiltonian (\ref{EffHamiltonian}) generates  a displacement of the
field but also induces a rotation on the qubit Bloch vector, the
simultaneity of both operations coming from the impossibility of
switching off the cavity-qubit interaction.

The measuring protocol consists of first encoding information
contained in the initial field state, $\hat{\rho}$, into the
CPB-qubit populations, which are then measured. The encoding process
is divided into four evolution steps: {\bf 1)} a coherent
displacement of the field; {\bf 2)} a $\pi/2$-pulse on the
CPB-qubit; {\bf 3)} a dispersive evolution without driving; and
finally {\bf 4)} another $\pi/2$-pulse on the CPB-qubit. The
displacement of the field as well as the pulses on the qubit are
driven by the external classical microwave field, whose complex
amplitude $\epsilon(t)=\epsilon$ and frequency $\omega_{d}$ are the
parameters under control.

In the first step we set $\omega_{d}=\omega-g^2/\Delta$ and
$\epsilon=\epsilon_{D}$, with $|\epsilon_{D}|\ll g+ \Delta^2/2g$.
This last condition implies, from (\ref{EffHamiltonian}) and
(\ref{n}), that we can neglect the components of the rotation axis
$\vec\Omega$ perpendicular to the $z$-direction. So, if initially
the qubit is in its ground state $|g\rangle$, it remains there, and
the evolution operator in a reference frame rotating with frequency
$\omega-g^2/\Delta$ is given by $\hat{U}_{\bf
1}(t_D,0)=\hat{D}(\alpha)$, where $\alpha=-i\epsilon_D t_D$, and
$t_{D}$ is the pulse duration.

In the second step we set the driving frequency
$\omega_{d}=\omega_{0}$ and
$\epsilon=|\epsilon_{\pi/2}|e^{i\phi_1}$, during a time
$t_{\pi/2}=\pi\Delta/4g|\epsilon_{\pi/2}|$. Then, according to
(\ref{n}), each Fock component of the state suffers  a
$\pi/2$-rotation of the Bloch vector about an axis whose
z-component, $(g^{2}/\Delta)(n+1/2)$, depends on the photon number
$n$. This component can be neglected under the condition
$|\epsilon_{\pi/2}|\gg g(\overline{n}+1/2)
\Leftrightarrow\overline{n}\ll |\epsilon_{\pi/2}|/g-1/2$, where
$\overline{n}$ is the mean number of photons in the state
\cite{nota}. In this case, the evolution operator is given by
$\hat{U}_{\bf 2}(t_D+t_{\pi/2},t_{D})= e^{i \Delta
\hat{a^\dagger}\hat{a}t_{\pi/2}}\hat{D}(\alpha_{\pi/2})
\hat{R}_{\vec{n}}(\pi/2)$ in the representation rotating with
frequency $\omega_{0}$ (which will be used from now on). This
essentially consists of a displacement, $\hat{D}(\alpha_{\pi/2})$,
of the field state by an amplitude $\alpha_{\pi/2}=-i\,e^{i\Delta
t_D}\int^{t_{\pi/2}+t_D}_{t_D} \epsilon_{\pi/2}e^{-i\Delta t}dt$,
and the rotation, $\hat{R}_{\vec{n}}(\pi/2)$,  of the qubit state by
an angle $\pi/2$ about an axis $\vec\Omega$ in the equatorial plane
of the Bloch sphere.

In the third step  we switch off the driving field, $\epsilon=0$,
and let the system evolve freely during a time
$t_{P}\equiv\pi\Delta/2g^{2}$ according to $\hat{U}_{\bf 3}(t_D
+t_{\pi/2}+t_{P},t_{D}+t_{\pi/2})= e^{i\frac{\pi\Delta^{2}}{2g^{2}}
\hat{a}^{\dagger}\hat{a}}e^{-i\frac{\pi}{2}
(\hat{a}^{\dagger}\hat{a}+\frac{1}{2})\hat{\sigma}_{z}}$. Because
the qubit is already in a superposition of the upper $|e\rangle$ and
lower $|g\rangle$ states, this is the step where the field and the
qubit get entangled, which is crucial for the transfer of
information from the field to the qubit. The fourth and last step in
the encoding process is another $\pi/2$-pulse on the qubit, with
$\epsilon\equiv|\epsilon_{\pi/2}|e^{i\phi_2}$, corresponding to an
evolution operator $\hat{U}_{\bf 4}$ analogous to $\hat{U}_{2}$.

Collecting all the steps of the encoding process we get the total
evolution $\hat{U}_T=\hat{U}_4\hat{U}_3\hat{U}_2\hat{U}\hat{U}_1$,
where $\hat{U}=e^{i(\Delta+g^2/\Delta)\big(\hat{a}^{\dagger}\hat{a}+
\hat{\sigma}_{z}/2\big)t_{D}}$ switches from a frame rotating with
frequency $\omega-g^2/\Delta$ to one rotating with frequency
$\omega_0$. Thus, $\langle\hat{\sigma}_z\rangle\equiv P_{e}-P_{g}
\equiv {\rm Tr}[\hat{\sigma}_z\hat{U}_T |g\rangle\langle g|
\hat{\rho}\,\hat{U}^{\dagger}_T]$ is straightforwardly calculated,
yielding
\begin{equation}
P_{e}-P_{g}=Tr\Big[\sin(\phi_1-\phi_2)\hat{\rho}\hat{D}(\beta)^{\dagger}
e^{-i\pi\hat{a}^{\dagger}\hat{a}}\hat{D}(\beta)\Big]\,,
\label{pe-pg}
\end{equation}
where $\beta\equiv \alpha+(2|\epsilon_{\pi/2}|/\Delta) \sin(
t_{\pi/2}\Delta/2) e^{-i\phi}$, with $\alpha=-i\epsilon_D t_D$ and
$\phi=(t_{D}+ t_{\pi/2}/2)\Delta+g^2t_D/\Delta +\pi/2-\phi_1$. When
$\phi_1-\phi_2=\pi/2$  we see from Eq.~(\ref{wigner}) that
$\langle\hat{\sigma}_z\rangle/\pi$ yields the value of the
initial-field Wigner function at the point $-\beta$. If the first
$\pi/2$-pulse is chosen so that $t_{\pi/2}\Delta/2=m\pi$, $m$
integer, then $\beta=\alpha$. By repeating the experiment for
different  $t_{D}$´s one can scan the whole phase space and thus
fully reconstruct the quantum state.
\begin{table}[b]
\begin{tabular}{|c|c|c|c|c|c|}
\hline
$\Delta$&$g$&$|\epsilon_D|$&$|\epsilon_{\pi/2}|$&$\kappa^{-1}$&$\gamma^{-1}$\\
\hline
$0.1$&$5\times 10^{-3}$&$0.025$&$0.025$&$160$ ns&$2\mu$s\\
\hline $0.3$&$5\times 10^{-3}$&$0.025$&$0.281$&$1000$ ns&$2\mu$s\\
\hline
\end{tabular}
\caption{ \emph{Experimental parameters}. $\Delta$, $g$,
$\epsilon_D$, and $\epsilon_{\pi/2}$ are expressed in units of the
transmission line frequency. The first set of parameters is reported
in \cite{blais:062320}; the second is the proposed one.
$\kappa^{-1}$ is the cavity lifetime and $\gamma^{-1}$ is the atom
lifetime.} \label{data}
\end{table}

\begin{figure}[t]
\begin{center}
    \scalebox{0.95}[0.95]{%
    \includegraphics*[angle=-90,width=8.5cm]{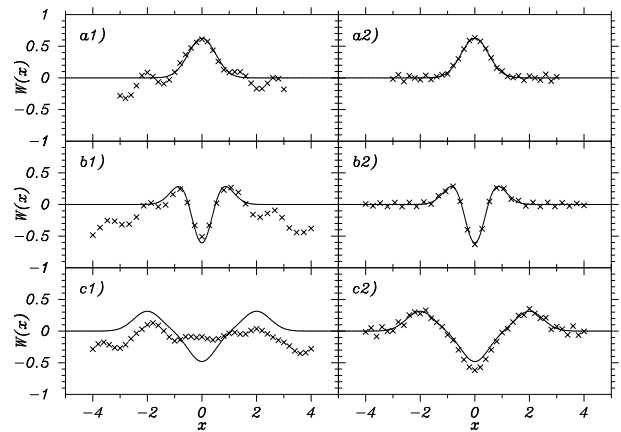}}
\end{center}
\caption{\footnotesize Wigner function $W(\alpha)$ on the real axis
$\alpha=x$ for different states of the cavity field: (a1) and (a2)
Vacuum state; (b1) and (b2) Fock state with $n=1$; (c1) and (c2)
Schr\"odinger-cat state
$\propto(|\alpha_0\rangle-|-\alpha_0\rangle)$, with $\alpha_0=2$.
The full line is the theoretical value and the crosses stand for the
values obtained by a numerical simulation of the measurement
protocol (see text for details). The numerical simulations on the
left-hand (right-hand) side were performed with the first (second)
row of parameters in Table \ref{data}. The transmission line
frequency is taken as $\omega=2\pi\times 10 \,GHz$.} \label{compare}
\end{figure}

Table \ref{data} displays a comparison between the parameters
reported in \cite{blais:062320} and the optimal parameters that we
propose. The performance of the protocol for these two sets of
parameters were tested by a numerical simulation, where each step of
the protocol was carried out evolving the system with the exact
JC-model plus the driving Hamiltonian. The Wigner function of the
field state was obtained using Eq.~(\ref{pe-pg}) where the
probabilities $P_e$ and $P_g$ were calculated for the final
entangled state of the system (see Fig.~\ref{compare}). With the
first set of parameters, the whole measurement protocol takes
approximately $100$ ns, which is less than the cavity lifetime
$\kappa^{-1}$ and much smaller than the atom lifetime $\gamma^{-1}$.
Moreover, the condition $|\epsilon_D|\ll g+\Delta^2/2g\approx
\Delta^2/2g=\omega$, used in the first step of the protocol, is well
satisfied. On the other hand, we see that the condition for the
$\pi/2$ rotation is not comfortably met, since it is not true that
$\overline{n}\ll|\epsilon_{\pi/2}|/g-1/2=4.5$ for the states of
interest. In fact, because the mean photon number increases after
the displacement of the field in the first step, {\it i.e.}
$\overline{n}\rightarrow \overline{n}+2\Re
e(\alpha^*\langle\hat{a}\rangle)+|\alpha|^2$, the conditions  for
the $\pi/2$ rotation and the dispersive regime approximation are
violated for greater values of $|\alpha|$ (see the numerical points
on the tails of the graphs  on the left in Fig.(\ref{compare})).
During the $\pi/2$ rotation of the CPB-qubit the field is also
displaced to $\beta=\alpha+(2|\epsilon_{\pi/2}|/\Delta) \sin(
t\Delta/2) e^{-i\phi}$, so the displacement $|\beta-\alpha|$ attains
its maximum value at the middle of the rotation, violating the
dispersive-regime approximation for the first set of parameters.
These considerations imply that the accuracy of the method is worse
for the tails of the Wigner functions, since probing them requires
larger displacements of the cavity field. The poor accuracy in
Fig.~2(c1) is due to the contribution of high-$n$ Fock states.
Finally, one should consider that with $\kappa^{-1}=160$ ns
decoherence effects may become appreciable at $100$ ns (duration of
the measurement protocol in this case) for fields with average
photon number larger than one. Thus, a higher-Q cavity should be
required.
%

With the detuning proposed in the second set of parameters the total
duration of the protocol would be about $300$ ns, for a cavity
lifetime $\kappa^{-1}\sim1000$ ns. The difference of one order of
magnitude between the values for the damping times in the two sets
of parameters in Table I is easily overcome with present technology.
Better lifetimes could be achieved, for example, by decreasing the
value of the capacitances $C_{0}$. On the other hand,
$|\epsilon_{\pi/2}|$ increases with the power of the external
radio-frequency source. The improvement in the reconstruction of the
Wigner function for this second set of parameters is displayed in
the graphs on the right-hand side of Fig.~\ref{compare}, for the
vacuum state, the Fock state with $n=1$, and a  Schr\"{o}dinger- cat
state. The main impact of using the new parameters is a great
reduction of the errors in the tails of the Wigner functions.
Decoherence, not taken into account in our simulations, would
further limit the maximum number of photons in the states
characterized. Temperature effects are negligible for typical
experimental values ($T=100$ mK  in \cite{wallraff:162}). Indeed,
for $T=100$ mK, and given that for our parameters we have
$\hbar\omega/k\approx480$ mK, the thermal occupation number is $\bar
n<0.008$. The reconstruction of states with more photons would
demand higher Q's, but the protocol would remain the same. We note
that the experiment could be used to continuously monitor the loss
of coherence of the field.

As for the measurement of the qubit population, a dispersive quantum
non-demolition scheme was carried out in \cite{wallraff:162}.
However, for the higher Q value considered here, this technique
would take a time of the order of the qubit lifetime. The direct
measurement of the qubit population could be accomplished in this
case by coupling to the qubit a single electron transistor (SET)
device, which is able to detect charge differences of the order of
$e$, as described, for example, in \cite{astafiev:180507}. The
influence of the SET on the qubit dynamics can be minimized by
turning the device on only at the moment of measurement, as
discussed in \cite{astafiev:180507}. The presence of an extra
superconducting lead connecting the SET to the qubit should not
increase significantly the decoherent effects on the cavity field
already introduced by the presence of the Cooper-pair box.

Experimental testing of this protocol would require the preparation
of simple field states. Coherent states are prepared by displacing
the initial state $|g,0\rangle$, which is accomplished by setting
the driving parameters $\omega_d=\omega-g^2/\Delta$ and
$\epsilon=\epsilon_D$, as in the first step of our protocol. This
will take a time $t_D=|\alpha|/|\epsilon_D|$ to be carried out,
where $\alpha$ is the amplitude of the coherent state.  The
generation of a Schr\"odinger cat-like state, {\it i.e.}
$(\ket{\alpha_0}\pm e^{i\varphi}\ket{-{\alpha_0}})/N$, where $N$ is
a normalization factor, is contained implicitly in the protocol
described in this paper, since after the whole evolution stage, and
before the qubit population is measured, the final entangled state
of the system is of the type
$(1/\sqrt{2})(|{\alpha_0}\rangle+\,e^{i\varphi}\,|-{\alpha_0}\rangle)
\otimes|e\rangle$
$+(1/\sqrt{2})(|{\alpha_0}\rangle-\,e^{i\varphi}\,|-{\alpha_0}\rangle)\otimes|g\rangle$.
Measuring the qubit population would project this state onto a
coherent superposition of two coherent states. The time required to
generate this state is the same as for our protocol.

Another example of interest is the one-photon Fock state: beginning
with the system in the state $\ket{g,0}$ one applies a $\pi$-pulse
on the qubit, setting $\omega_d=\omega_0$  and
$\epsilon=\epsilon_{\pi}$ during a time
$t_{\pi}=\pi\Delta/2g|\epsilon_{\pi}|$. Choosing $t_\pi\Delta/2 = m
\pi$, with $m$ integer, the state $\ket{e,0}$ is prepared. Then, we
tune the qubit frequency $\omega_0$ into resonance with the cavity
mode (by changing the magnetic flux $\Phi$) and let the system
complete a Rabi oscillation ($t_{Ra}=\pi/2g$), so the resulting
state is $\ket{g,1}$. Next we change the magnetic flux again to take
the system back to the dispersive regime. This method would require
rapid switching of the flux (less than 1 ns), a challenge for
present experiments. Alternatively, time-dependent magnetic fluxes
could be used to tune the qubits into and out of resonance with the
cavity field, as recently suggested in \cite{nori2}. For the second
set of parameters in Table \ref{data} and $|\epsilon_{\pi}| =
0.3\,\omega$ ($m=15$) the total time for this process would then be
$t_{Ra}+t_\pi\sim 10$ ns.


In conclusion, we propose here an experiment to completely
characterize the electromagnetic field of a quasi 1-D
superconducting transmission line resonator, which is always
interacting with a CPB-qubit, by directly measuring its Wigner
function. This is carried out by measuring the CPB-qubit population
after the application of a series of driving pulses, induced by an
external microwave field coupled to the resonator. Our numerical
simulations, for realistic parameters, show that this method is
within reach of present experimental setups.

We thank F. Schackert and C. H. Lewenkopf for useful discussions and
CAPES, CNPq, FAPERJ, and the Millennium Institute for Quantum
Information for support.


\begin{thebibliography}{10}
\bibitem{nota}
Strictly speaking, we should impose this condition to every photon
number $n$ present in the state; or, alternatively, ask that the
distribution in photon number be very localized around
$\overline{n}$.
\end{thebibliography}
\end{document}